\documentclass[aps,pra,superscriptaddress,nofootinbib,twocolumn]{revtex4-2}

\usepackage{graphicx}
\usepackage{amsmath}
\usepackage{physics}
\usepackage{caption}

\newcommand{\um}{\mu \mathrm{m}}
\newcommand{\nm}{\mathrm{nm}}
\newcommand{\eps}{\varepsilon}

\newcommand{\micron}{\mu \mathrm{m}}
\newcommand{\mm}{\mathrm{mm}}
\newcommand{\fp}{Fabry-P\'erot}
\newcommand{\re}{\mathrm{Re}}
\newcommand{\im}{\mathrm{Im}}
\newcommand{\out}{\mathrm{out}}

\begin{document}
\title{Coherent absorption reveals colors hidden in a grey Fabry-Pérot cavity}

\author{Giuseppe E. Lio}
\thanks{These authors contributed equally to this work.}
\affiliation{CNR-Istituto Nanoscienze, Laboratorio NEST, Piazza San Silvestro 12, 56127 Pisa, Italy}

\author{Giulio Carotta}
\thanks{These authors contributed equally to this work.}
\affiliation{Dipartimento di Fisica “E. Fermi”, Università di Pisa, Largo Pontecorvo 3, 56127 Pisa, Italy}

\author{Lorenzo Lavista}
\affiliation{Dipartimento di Fisica “E. Fermi”, Università di Pisa and CNR-Istituto Nanoscienze, Largo Pontecorvo 3, 56127 Pisa, Italy}

\author{Andrea Camposeo}
\affiliation{CNR-Istituto Nanoscienze, Laboratorio NEST, Piazza San Silvestro 12, 56127 Pisa, Italy}

\author{Giacomo Venturi}
\affiliation{Dipartimento di Fisica “E. Fermi”, Università di Pisa, Largo Pontecorvo 3, 56127 Pisa, Italy}

\author{Agnese Guernieri}
\affiliation{Dipartimento di Fisica “E. Fermi”, Università di Pisa, Largo Pontecorvo 3, 56127 Pisa, Italy}

\author{Alessandro Pitanti}
\affiliation{Dipartimento di Fisica “E. Fermi”, Università di Pisa and CNR-Istituto Nanoscienze, Largo Pontecorvo 3, 56127 Pisa, Italy}

\author{Simon A. R. Horsley}
\affiliation{School of Physics and Astronomy, University of Exeter, Stocker Road, Devon, EX4 4QL, UK}

\author{Giuseppe C. La Rocca}
\affiliation{NEST, Scuola Normale Superiore, Piazza dei Cavalieri 7, I-56126 Pisa, Italy}

\author{Alessandro Tredicucci}
\affiliation{Dipartimento di Fisica “E. Fermi”, Università di Pisa and CNR-Istituto Nanoscienze, Largo Pontecorvo 3, 56127 Pisa, Italy}

\author{Simone Zanotto}
\email{simone.zanotto@nano.cnr.it}
\affiliation{CNR-Istituto Nanoscienze, Laboratorio NEST, Piazza San Silvestro 12, 56127 Pisa, Italy}

\date{\today}

\begin{abstract}
Thin dielectric films are known to show distinct colors, responsible for the iridescence of various natural and artificial objects such as insect wings and soap bubbles. In the present article we show that a specialized thin film Fabry-Pérot resonator, that we name conductor-dielectric-conductor (CDC) matched cavity, appears instead completely grey when observed under ordinary conditions (i.e.~by analyzing the transmitted or reflected incoherent white light). Nonetheless, the matched CDC cavity still retains spectral information, that clearly appear when the cavity is analyzed by the coherent absorption technique. The CDC system is simply a dielectric thin film sandwiched by two conducting interfaces, whose conductivity shall be appropriately matched depending upon the dielectric refractive index. In practical applications, the ideal conducting interfaces can be safely replaced by thin metal films, making the system easily applicable in particular for cryptographic purposes. We indeed demonstrated experimentally that a visually recognizable thin-film color pattern can be concealed to an ordinary observer, and finally recovered through a dedicated coherent absorption decoding apparatus.
\end{abstract}

\maketitle

\section{Introduction} Linear wave physics, and in particular the propagation of electromagnetic waves in time-invariant local media, is far from being an exhausted field of study. In this framework, a huge source of complexity and richness spawns from the strongly nonlinear mapping between the set of spatial permittivity distributions and the space of the corresponding Maxwell equation solutions \cite{eliezer23}. In other words, the inverse scattering problem still poses significant theoretical and numerical challenges \cite{hopcraft}. Focusing on the applications, such complexity reverberates in the flourishing scientific production about intricated, yet precisely functional, two- or three-dimensional photonic structures created by inverse design \cite{molesky18}, or by advanced techniques that leverage specific classes of disorder to tackle open problems in radiofrequency engineering and in integrated optics \cite{yu21, cao22apr}. It is instead quite surprising that even in the restricted space of one-dimensional structures, i.e.~multilayered planar systems, new physical phenomena are being discovered only recently.  These include the Fano resonant optical coating \cite{elkabbash21}, ultrabroadband near-perfect absorbers \cite{kim25}, perfect absorption in ultrathin, highly lossy planar films \cite{kats16, sakotic23}, passive radiative coolers \cite{raman14, lee23}, topological darkness \cite{cusworth23} and hyperbolic metamaterials \cite{guo20}. It is thus possible, with relatively simple fabrication techniques that do not involve any lithographic patterning, to implement tailored angular and spectral responses in reflection, transmission, absorption, and emission. 

Within linear optical physics, however, the full potential of a system is revealed when moving from ordinary one-beam measurements to structured illumination studies, where many input channels are simultaneously populated with wavefronts having precise phase and amplitude relations \cite{bender22}. From the theoretical point of view, this approach allows to natively probe the singular value decomposition of the system's scattering matrix $S$, and its possibly nontrivial topology \cite{guo23prb, guo23prl}. In a multilayer made of isotropic materials, there are only two input and two output channels (the beams incoming/outgoing from the two sides of the film, with a fixed linear polarization), the $S$ matrix is two-by-two, and there are two singular values/singular vector pairs. By injecting into the system a two-beam field that matches a singular vector, the absorbed energy is dictated by the corresponding singular value, and is in general different from the single-beam absorptance. 
If a singular value is zero, the so-called coherent perfect absorption (CPA) is attained, with 100\% absorption from an otherwise semitransparent film \cite{chong10, wan11, baranov17}. CPA has been demonstrated in a variety of systems, mostly in photonics, with some outstanding examples in acoustics \cite{farooqui22}, in coherent matter waves \cite{muellers18}, and in thermal transport \cite{li2022}. In any case, CPA usually occurs in correspondence to a specific resonance, either broadband or narrowband, where a generalized critical coupling condition is satisfied \cite{zanotto14, zanotto16, sweeney20}. Besides its effects on coherent absorption, the presence of a resonance is normally observable also in ordinary absorption, in transmission, and in reflection - whose spectra are often Lorentzian or Fano lineshapes centered around the involved frequency. 

\begin{figure*}
\centering
\includegraphics[width = \textwidth]{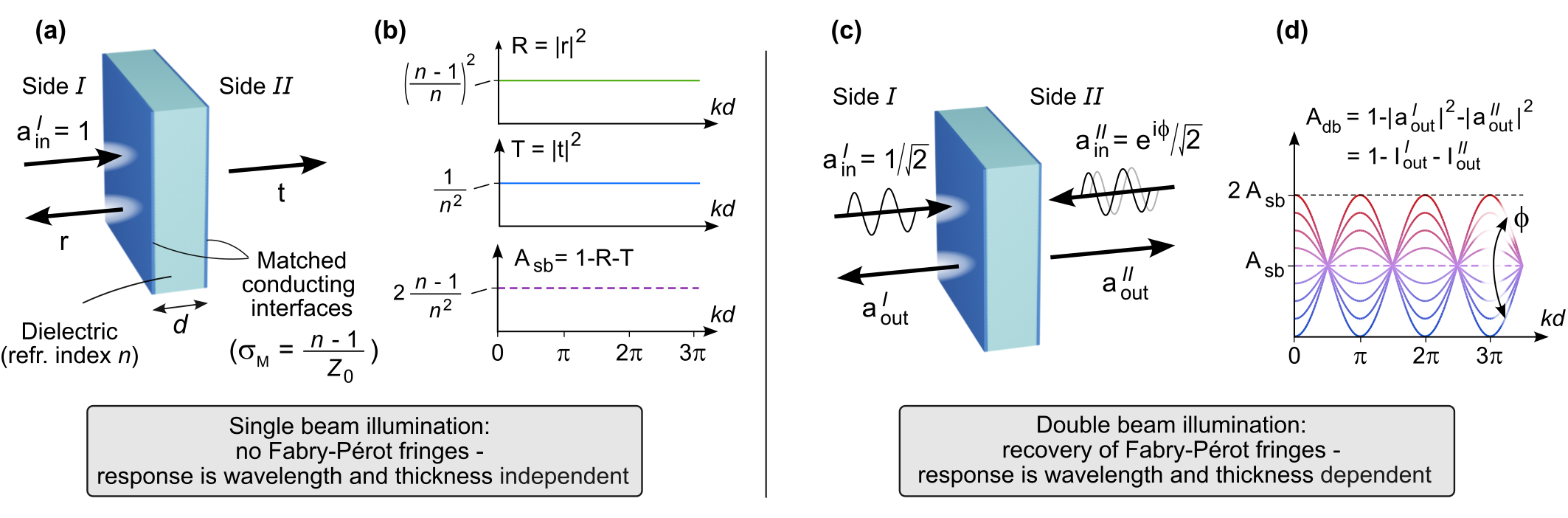}
\captionsetup{margin=10pt,font=small,labelfont=bf,justification=raggedright}
\caption{\textbf{Concept of the conductor-dielectric-conductor (CDC) grey cavity - } A lossless dielectric layer sandwiched between conducting interfaces having a matched conductivity (a) exhibits flat single-beam reflectance $R$, transmittance $T$ and absorptance $A_{sb}$ spectra. Here, with ``flatness'' we mean that the ordinary spectra are independent of either wavelength (entering the wavevector $k = 2\pi n/
\lambda_0$) and thickness $d$. When considering a double-beam experiment, where coherent light is impinging on both sides of the cavity (c), the double-beam absorptance $A_{db}$ is modulated by the input beams' relative phase $\phi$ (d). Importantly, such modulation can be zero (for $kd = (2m+1)\pi/2$, with $m$ integer), or maximum (for $kd = m\pi$). Such maximum can reach unity when $2 A_{sb} = 1$, i.e.~when $n = 2$ (see also panel (b)).}
\end{figure*}
In the present paper we radically challenge this common acquisition, demonstrating that it is possible to design a special \fp\ cavity that has \textit{spectrally flat} reflectance and transmittance (and hence also flat single-beam absorptance via the ordinary relation $A_{sb} = 1-R-T$), but \textit{spectrally modulated} double-beam absorptance. The operation principle is based on the behavior of a matched thin conductor as a broadband and lossy antireflection coating. For an external observer, our cavity is \textit{grey} under ordinary, single-beam observations, and \textit{colored} under coherent illumination. Structural color in thin films are known since long time, also for their potential in anticounterfeting and information encryption \cite{xuan21, dobrowolski89}. However, known approaches relied on decoding using ordinary reflected or transmitted light. We will eventually show that the proposed special \fp\ cavity can be used to encrypt structural-color based information with an additional layer of security, since the concealed information can be retrieved only through a coherent absorption measurement.

\section{Results} 
The special \fp\ cavity under study is sketched in Fig.~1a: a lossless dielectric layer of thickness $d$ and refractive index $n$ is sandwiched between mirrors constituted by optically thin, two-dimensional conducting sheets with surface conductance $\sigma$, being in general a complex value. When the mirrors' conductance matches the special, real value given by 
\[
\sigma_M = \frac{n-1}{Z_0},
\]
where $Z_0 = 377\, \Omega$ is the vacuum impedance, the reflectance and transmittance $R$ and $T$ are spectrally flat, implying also a flat single-beam absorptance $A_{sb}$ (Fig.~1b). On the contrary, when the system is illuminated from the two sides by beams having equal amplitude and variable relative phase $\phi$, the double-beam absorptance $A_{db}$ acquires a dependence on both $\phi$ and wavelength (Fig.~1c-d). More precisely, one has
\begin{equation}
\label{eq:Adb}
A_{db} = A_{sb} \left( 1 - \cos(kd)\cos(\phi) \right)
\end{equation}
where $k = 2\pi n /\lambda_0$, being $\lambda_0$ the vacuum wavelength. 
 
\begin{figure*}
\centering
\includegraphics[width = \textwidth]{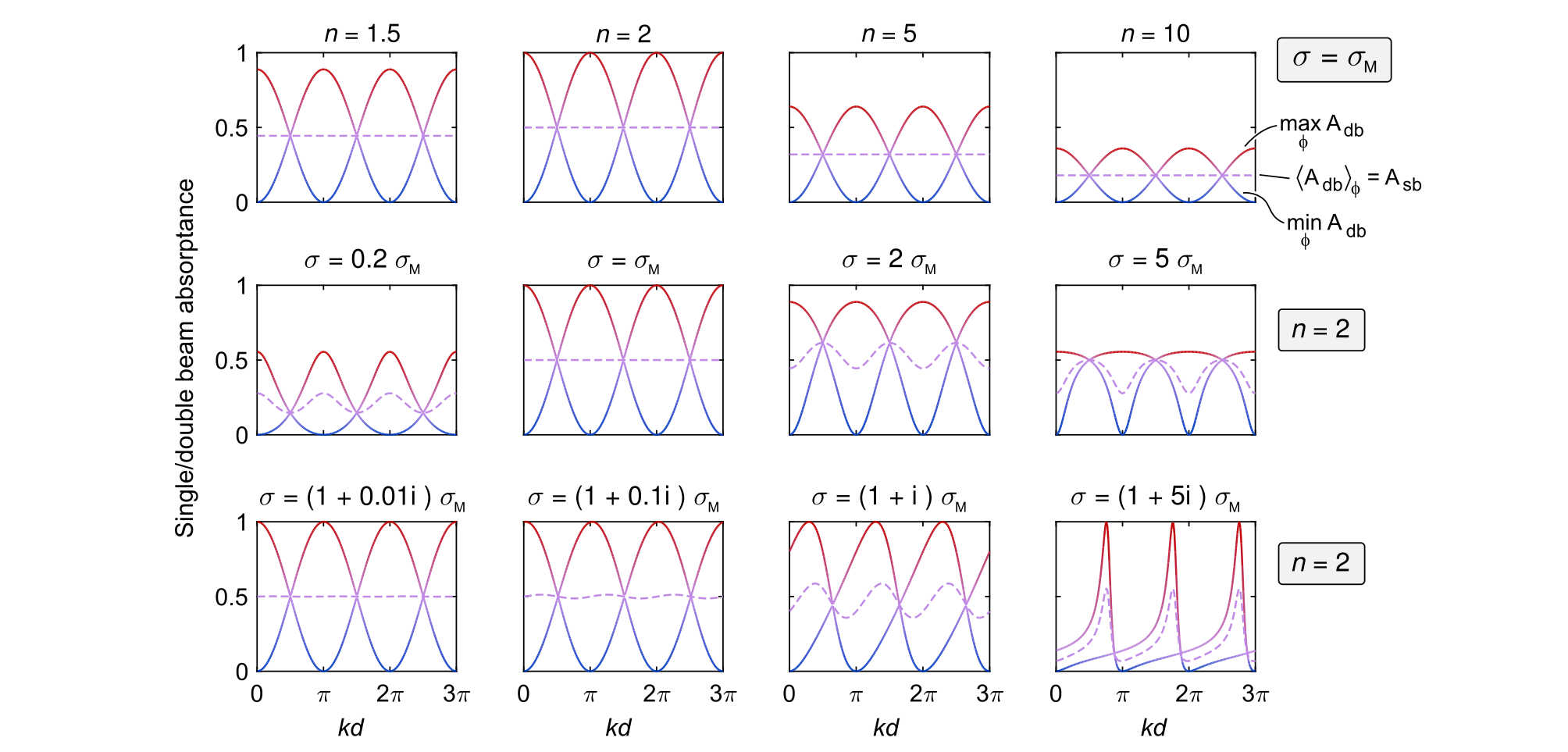}
\captionsetup{margin=10pt,font=small,labelfont=bf,justification=raggedright}
\caption{\textbf{Single and double beam absorptance of matched and mismatched CDC cavities -} Maximum, minimum, and average double-beam spectral absorptance, calculated with respect to the input dephasing $\phi$, for several combinations of dielectric refractive index $n$ and interface conductivity $\sigma$. The average double-beam absorptance always coincides with the single-beam absorptance. \textit{Top row}: spectra for the matched cavity ($\sigma = \sigma_M$), for different values of the dielectric refractive index. In all cases $\max\,  / \min\,  A_{db}$ are sinusoids and $\left\langle A_{db} \right\rangle = A_{sb}$ is constant. The case $\sigma = \sigma_M$, $n = 2$ represents a special case, where $\max\, A_{db}$ can reach unity. \textit{Middle row}: absorptance spectra for $n = 2$ and variable, real $\sigma$. In general, $\max\,  / \min\,  A_{db}$ and $\left\langle A_{db} \right\rangle = A_{sb}$ are non-sinusoidal, oscillating functions. \textit{Bottom row}: absorptance spectra for $n = 2$ and variable $\sigma$ where a spurious imaginary component is included. In general, $\max\,  / \min\,  A_{db}$ and $\left\langle A_{db} \right\rangle = A_{sb}$ are non-sinusoidal, oscillating functions, with warped lineshapes.}
\end{figure*}
Such unique behavior can be described analytically by solving the field propagation in the conductor-dielectric-conductor (``CDC'') geometry and inspecting the resulting formulas. Following the procedure detailed in Suppl.~Sect.~I, the cavity reflection and transmission coefficients are 

\begin{equation}
\label{eq:r}
r=-\frac{\alpha_{+-}\, \alpha_{--}\, e^{ikd}- \alpha_{-+}\, \alpha_{++}\, e^{-ikd}}{\alpha_{--}^2\, e^{ikd}-\alpha_{++}^2\,e^{-ikd}},
\end{equation}
\begin{equation}
\label{eq:t}
t=\frac{\alpha_{-+}\, \alpha_{--} - \alpha_{+-}\, \alpha_{++}}{\alpha_{--}^2\, e^{ikd}-\alpha_{++}^2\,e^{-ikd}},
\end{equation}
where
\[
\alpha_{++} = n+1+\sigma Z_0, \qquad \alpha_{+-} = n+1-\sigma Z_0,
\]
\[
\alpha_{-+} = n-1+\sigma Z_0, \qquad \alpha_{--} = n-1-\sigma Z_0.
\]

Neglecting the material dispersion, that can be mitigated by a proper material choice, the strongest contribution to the wavelength dependence in Eqs.~(\ref{eq:r}-\ref{eq:t}) is here dictated by the phase factor $e^{ikd}$. Such wavelength dependence survives in general also after taking the squared moduli $|r|^2$ and $|t|^2$, i.e.~when considering the single-beam spectroscopic observables. However, if the factor $\alpha_{++}$ is zeroed, which happens for the aforementioned choice of $\sigma = \sigma_M$, drastically simplified formulas emerge:
\[
r=\frac{1-n}{n}, \qquad t=\frac{e^{ikd}}{n}.
\]
The immediate consequence is that the single-beam spectra lose any dependence on $kd$. However, a subtler effect happens as long as double-beam absorptance is considered: here, indeed, the phase factor $e^{ikd}$ that has survived inside the $t$ coefficient plays a crucial role and manifests ultimately in the $A_{db}$ formula expressed in (\ref{eq:Adb}), derived in Suppl.~Sect.~II and III. Another view on the phenomenon can be gained by noticing that the matched conducting interface acts as a \textit{wavelength-independent antireflection coating} for waves that travel from the interior of the dielectric material towards the exterior (see Suppl.~Sect.~IV). Hence, for single-beam illumination of the system (say from the left side) the dielectric region is only populated with rightward-propagating waves, and \fp\ interferences do not occur. Similarly, upon incidence from the right side of the system, the dielectric is populated solely with leftward-propagating waves. However, for double-beam illumination, the superposition principle dictates now that a standing wave is formed within the cavity, leading to an absorption that depends on the field value in correspondence of the absorbing interfaces.

A further inspection of the formulas indicates that an even more special condition may occur. The key is the formula for single-beam absorption, that for the $\sigma = \sigma_M$ case reads $A_{sb} = 2 (n-1)/n^2$. Here, if $n = 2$, one gets $A_{sb} = 1/2$, which in turn guarantees that $A_{db}$ has a full excursion in the [0,1] interval when $kd = m\pi$, with $m$ an integer. Such exceptional condition is related to the maximised absorption occurring in a free-standing conducting interface with $\sigma = Z_0/2$, or similarly in an ultrathin, epsilon-near-zero material at the Woltersdorff thickness \cite{pu12, radi15, li15prb}.
Yet, the phenomenon in our CDC system is much more powerful. Indeed, a free-standing conducting interface with $\sigma = Z_0/2$ or a thin film at the Woltersdorff thickness has spectrally flat 50\% absorption, but also spectrally flat 0-100\% \textit{coherent} absorption, since the subwavelength nature of these systems cannot provide any resonant - and hence wavelength-dependent - phenomena to the coherent absorption. 

\begin{figure*}
\centering
\includegraphics[width = \textwidth]{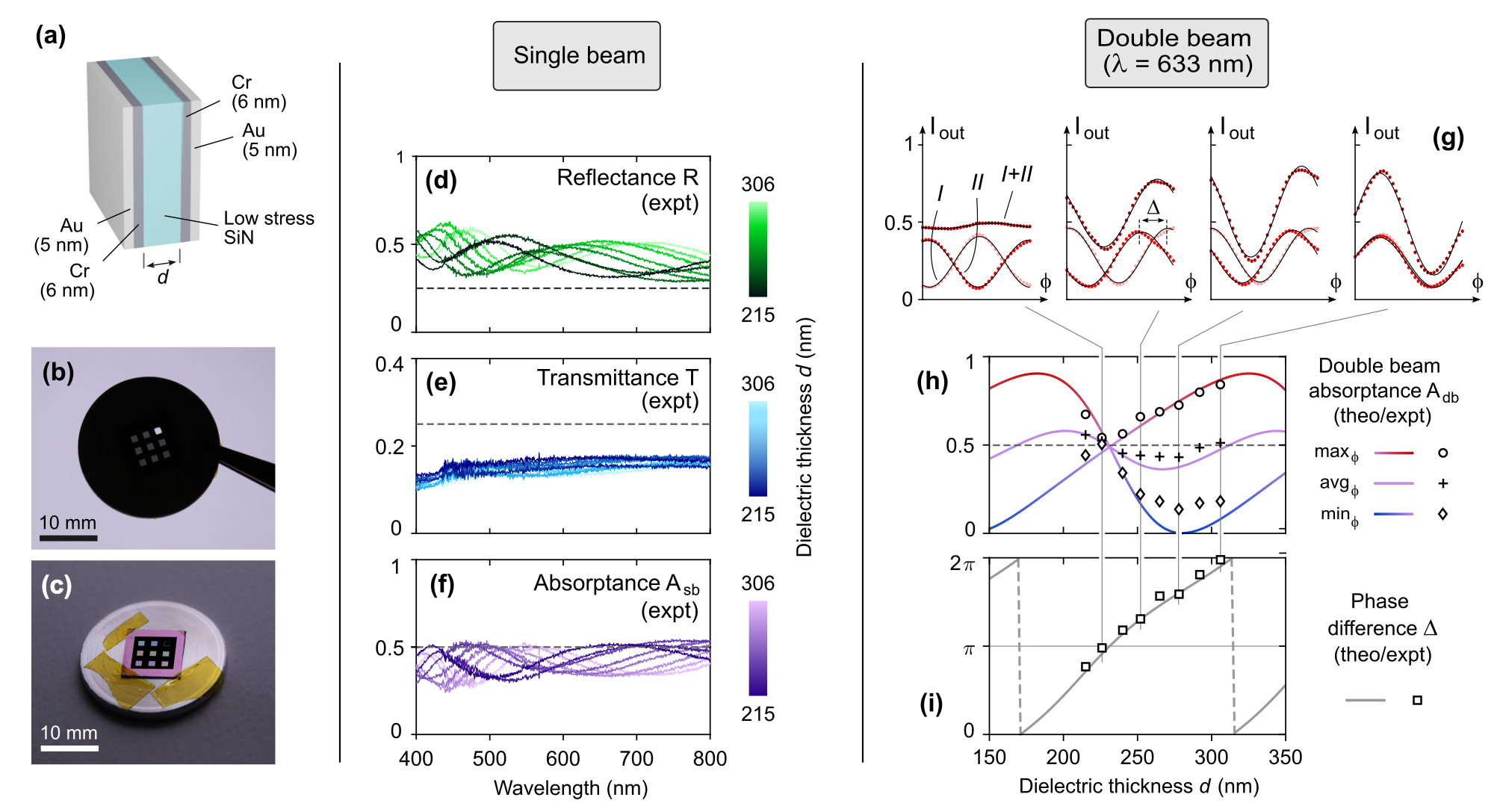}
\captionsetup{margin=10pt,font=small,labelfont=bf,justification=raggedright}
\caption{\textbf{Single- and double-beam measurements on variable-thickness CDC cavities -} (a) Schematic of the investigated samples, that are conductor-dielectric-conductor cavities where the ideal, zero-thickness conductor is implemented by a thin metallic bilayer. (b) Photograph of an array of CDC membrane samples having different dielectric thickness, backside-illuminated with white light. All membranes appear identical (the top-right membrane is missing, leaving a hole as a blank reference). (c) Photograph of the same CDC membrane array, illuminated from the frontside with white light. The CDC samples show weakly saturated colors, with variable hues. A black frame is artificially introduced to better recognize the intrinsic colors, see also Extended Data Fig.~1. (d)-(f) Reflectance, transmittance, and single-beam (i.e.~ordinary) absorptance spectra collected from the CDC sample array shown in (b)-(c). The quantitative spectra confirm the qualitative perceptual colors observed in the photographs. (g) Output light intensity observed when samples of different thickness are illuminated from two sides with coherent, dephased beams (as in Fig.~1c). Traces $I$ and $II$ refer, respectively, to output light intensity $I_{\out}^{I}$ and $I_{\out}^{II}$, detected on either sides of the membranes (cfr.~Fig.~1c). The total output intensity ($I+II$) can be weakly or strongly modulated, depending on the dielectric thickness. (h) Maximum, average, and minimum (with respect to the dephasing $\phi$) double-beam absorptance as a function of the dielectric thickness.  (i) Dependence of the phase $\Delta$ (see (g) for the definition) upon the dielectric thickness. In (h)-(i), points are experimental data and lines are full-wave simulations.}
\end{figure*}
The conditions laid out in the above are extremely robust, also thanks to their analytical simplicity; yet a caveat must be highlighted. Indeed, the stronger condition where $n = 2$ and $\sigma = 1/Z_0$ is hard to achieve due to the inevitable intrinsic dispersion of dielectrics and conductors. Even the weaker condition where the dielectric permittivity is free ($n \neq 2$), the conductor matching requirement $\sigma = \sigma_M = (n-1)/Z_0$ poses the challenge of finding a material with purely real conductivity with the correct wavelength dependence. As this is hard to achieve in reality, we shall now consider the effect of phenomenological deviations of $n$ and $\sigma$ from the matched conditions; further details on real materials will be given later, when dealing with an experimental realization. For the present analysis, we consider in Figure 2 the minimum, average and maximum (with respect to $\phi$) double-beam absorption spectra. 
The average $A_{db}$ spectra can be shown to be equal to $A_{sb}$, while the maximum/minimum $A_{db}$, that are the singular values of $S$, can be derived analytically (see Suppl.~Sect.~II). Notice that, if $\sigma \neq \sigma_M$, Eq.~(\ref{eq:Adb}) no longer holds. The top row of Fig.~2 illustrates the effect of a variable $n$, while keeping the conductivity always at the corresponding matched condition. As anticipated, $A_{sb}$ is always flat, while $A_{db}$ is wavelength-dependent following Eq.~(\ref{eq:Adb}). Notice that $\min A_{db}$ has always zeros for $kd = m\pi$, while $\max A_{db}$ is in general smaller than 1, except for the double-matching case $n = 2, \sigma = \sigma_M$. The intermediate row of Fig.~2 illustrates the effect of a mismatched, yet purely real, $\sigma$. Here, $A_{sb}$ is not flat, $\min A_{db}$ still has zeros at $kd = m\pi$, and $\max A_{db} < 1$ in general. Moreover, the functional dependence upon $kd$ is no longer the sinusoid of Eq.~(\ref{eq:Adb}), rather it is a kind of higher-finesse \fp -like lineshape. Finally, the lower row of Fig.~2 shows what happens if $\sigma$ acquires a spurious imaginary component: $A_{sb}$ is not flat, the lineshapes are warped, $\min A_{db}$ ($\max A_{db}$) have zeros (ones) at $kd = m\pi$.

Supported by the theory, we moved hence to an experimental demonstration, identifying a material system that closely satisfies the double-matching condition (Fig.~3a). As a dielectric material, the choice has naturally fallen on silicon nitride (SiN), as its refractive index is close to 2 in the visible and near-infrared spectral range. Moreover, employing low-stress, non-stoichiometric SiN eases the fabrication of rather large ($1~\mm^2$) free-standing membranes. While in principle one can use two-dimensional materials such as graphene - that indeed has a tunable, almost purely real conductivity in the visible and near-infrared \cite{stauber08,pellegrino10} - we decided to rely on a metal film manufacturable through thermal evaporation, in view of a simpler process and easier scalability. Being a bulk material, a metal is described by its relative permittivity $\eps_m$, however, a metal thin film can be safely approximated as a conducting interface with $\sigma = -i d_m (\eps_m-1) \eps_0 2 \pi c/\lambda_0$, where $d_m$ is the metal thickness, $\eps_0$ is the vacuum permittivity, $c$ is the speed of light, and $\lambda_0$ the free-space wavelength. Looking for a real $\sigma$, one needs $\re(\eps_m)$ as close as possible to 1; then, the required $\sigma$ can be retrieved by tuning $d_m$. (More precisely, what matters is to minimize the ratio $(\re(\eps_m)-1)/(\im{\eps_m})$). Among the common metals, chromium satisfies the aforementioned condition rather well in a quite broadband region, with an optimal thickness of $4\ \nm$. However, to avoid the risk of islanding effects, we increased the thickness to $6\ \nm$; moreover, to avoid surface oxidation, we added a $5\ \nm$ gold layer. The deviation of $\sigma$ from $\sigma_M$ for such metal films are discussed in Suppl.~Sect.~V.

\begin{figure*}
\centering
\includegraphics[width = \textwidth]{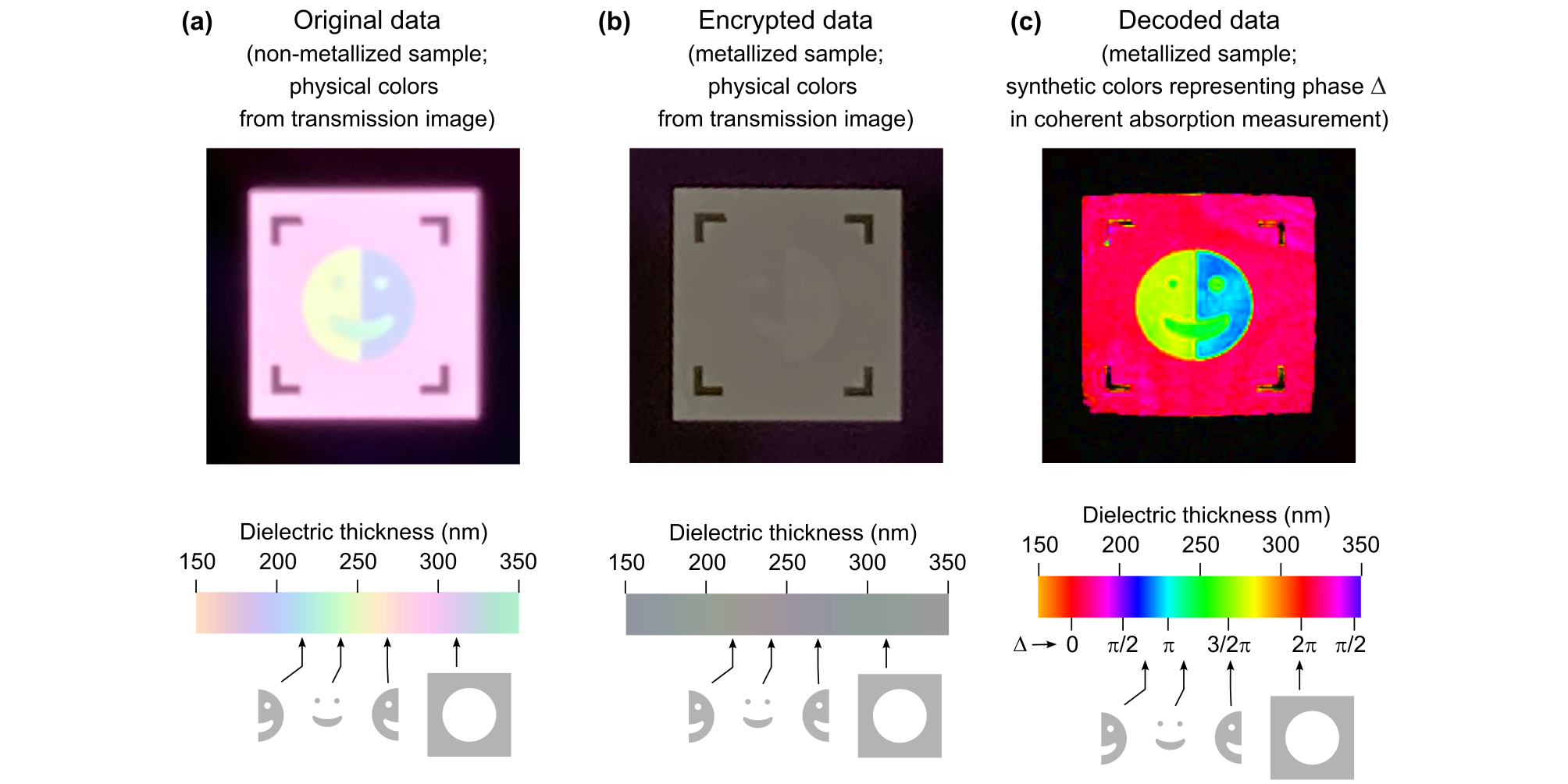}
\captionsetup{margin=10pt,font=small,labelfont=bf,justification=raggedright}
\caption{\textbf{Cryptography based on the colors hidden in CDC grey cavities -} (a) Transmission photograph of a silicon nitride membrane where different regions have different thicknesses, resulting in a ``smiley'' image caused by interference colors. (b) Transmission photograph of the same membrane, coated on both sides with the metallic bilayer of Fig.~3a, resulting in CDC grey cavities where the structural colors are strongly suppressed. The colorbars of (a) and (b) are obtained by converting the simulated transmittance spectra to rgb values using the CIE1931 model with D65 illuminant and gamma correction. (c) Image of the ``smiley'' reconstructed through coherent absorption at fixed wavelength ($633\ \nm$). Here, the colors represent the phase $\Delta$ measured locally with the approach of Fig.~1c and Fig.~3g-i. The synthetic colors of (c) resemble the hue of the natural colors of (a), despite the metallized CDC sample being grey in ordinary, single-beam transmission. The size of the imaging window is $5 \times 5\ \mm$.}
\end{figure*}
In principle, to demonstrate the ``colored CPA'', one needs to probe a sample with coherent light of different wavelengths. As this would require either several lasers or a widely tunable one, we tackled the issue resorting to the  equivalent effects of wavelength and thickness tuning, stemming from the $kd$ dependence in Eq.~(\ref{eq:Adb}). We hence fabricated a set of 8 membranes, with SiN thickness ranging from $215$ to $306\ \nm$, on a single chip whose fabrication details are given in Methods. Figures 3b and 3c report photographs of the sample, taken respectively with backside illumination or frontside illumination. In either cases light is white and diffuse. It can be clearly noticed that in Fig.~3b all the membranes have an almost indistinguishable shade of grey, indicating that the spectral transmittances of all the cavities are flat and equal. In Fig.~3c the membranes appear instead slightly colored, an effect originating from the adoption of a real metal bilayer instead of the ideal matching interface. These visual clues are fully confirmed by spectroscopic white-light reflectance and transmittance measurements, reported in Fig.~3d-e. Reflectance spectra have a baseline of $\approx 40\%$ with $\approx 25\%$ wide \fp -like fringes; transmittance spectra are instead exceptionally flat, with a baseline of about $17\%$. In Fig.~3d-e we report as a dashed line the ideal reflectance and transmittance values expected for a cavity with $n = 2$ and $\sigma = \sigma_M$: $R = T = 25\%$. Fig.~3f reports the single beam absorptance, derived from $R$ and $T$.

We then performed coherent absorption measurements at $633\ \nm$ on the various membranes, using the setup detailed in Methods that implements the concept in Fig.~1c. While sweeping the relative input phase $\phi$, we monitored the output intensities $I_{\out}^{I,II} = |a_{\out}^{I,II}|^2$; the overall output intensity is then determined as $I_{\out}^{I}+I_{\out}^{II}$. It can be seen from Fig.~3g that $I_{\out}^{I}+I_{\out}^{II}$ has a sinusoidal behavior, with amplitude dependent on the dielectric thickness. Since $A_{db} = 1-I_{\out}^{I}-I_{\out}^{II}$, this proves that the double beam absorptance can be either strongly modulated or almost constant upon a sweep of $\phi$ (Fig.~3h). Noticeably, this effect stems from the different dephasing $\Delta$ that occurs among $I_{\out}^{I}$ and $I_{\out}^{II}$, as sketched in Fig.~3g. On the other hand, the sinusoids described by $I_{\out}^{I,II}$ as a function of $\phi$ have always the same amplitude, for all values of $d$ (This feature is general for any symmetric coherent absorbing system illuminated with equal-intensity beams, see Suppl.~Sect.~II and VI). From Figs.~3d-i it stands out that an easily fabricated sample implements with a good fidelity the ideal features of the matched cavity, despite some spectral ripples that appear in the $R$ and in the $A_{sb}$ spectra. Nonetheless, $A_{db}$ has a very large modulation range, with $\min A_{db} - \max A_{db}$ spanning from zero to almost $90\%$. Also the phase difference $\Delta$ is almost linearly dependent on $d$, as predicted by the model for the perfectly matched interfaces (where $\Delta = 2kd$). The modeling accuracy gets excellent when comparing the experimental data with the solid lines in Fig.~3h-i, that represent simulations performed assuming finite-thickness metal layers and true material dispersion. Besides CPA-related quantities, also the complete theoretical modeling of single-beam spectra achieves an outstanding level of precision (see Suppl.~Sect.~VII).

Leveraging on this novel and robust physical phenomenon, we devised an experiment to demonstrate the cryptograhic potential of colored CPA in a grey cavity. For this purpose, we fabricated a 5 mm wide sample in which the thickness of the SiN layer varies spatially according to a “smiley” image. Fig.~4a reports a photographic image of the light transmitted from the sample before metal deposition, i.e.~the variable thickness, bare SiN membrane. Here, the colors stem from ordinary thin-film interference; as the SiN/air interface has a moderate refractive index contrast of approximately 2 to 1, the colors have intermediate saturation and rather high luminosity. The situation changes drastically after depositing the Cr/Au bilayer on both membrane sides. Now, the image of Fig.~4b is observed: as expected, all the regions are characterized by an almost fully unsaturated, moderate-luminosity grey shade, according to the theoretical prediction. However, the information physically encoded in the sample is not washed out, and can be accessed at any time through an interferometric measurement. Relying on a slightly modified CPA setup, where the detectors are replaced by a CCD camera, and relying on a specialized data analysis protocol (Methods and Suppl.~Sect.~VIII), all the CPA-related quantities can be accessed pixel-by-pixel. In particular, it is possible to retrieve the phase $\Delta$, that is directly connected to the dielectric thickness $d$. The result is shown in Fig.~4c, where we encoded the values of $\Delta$ in \textit{false colors}. We used a fully saturated cyclic hue color scale, that naturally maps to a physically cyclic quantity such as a phase (see also Fig.~3i). Noticeably, when the reference level for this synthetic color scale is properly set, it matches quite well the sequence of hues observed in the physical colors of Fig.~4a. In other words, with a coherent measurement performed with monochromatic light (here, $633\ \nm$, but any wavelength can be used in principle), it is possible to retrieve the visual appearance of the original sample, that was permanently concealed from any single-beam measurement. Due to the dramatically different effort needed to collect ordinary transmission images and interferometric data - Figs.~4a-b have been collected with a low-end smartphone, Fig.~4c with a dedicated laboratory apparatus - we believe that the presented method can be fruitfully employed in a cryptographic scheme where the secret can be decrypted only by a restricted number of recipients. 

\section*{Discussion and conclusions}
In our work we proposed a special \fp\ cavity, based on a matched conductor-dielectric-conductor geometry, that shows spectrally flat single-beam spectroscopic observables (i.e.~reflectance, transmittance, and absorptance). Nonetheless, the cavity hides a color information that can be unveiled by a coherent absorption measurement; the system can hence pose the basis for an information encryption protocol. A pioneering study on cryptography based on coherent absorption has indeed been reported \cite{xomalis19}, however, that work dealt with a fiber-based system that lacks direct visual access to the uncrypted information. On the contrary, a planar, thin-film approach, possibly supported in the future by advanced technologies such as two-dimensional materials \cite{li25prb}, metasurfaces \cite{alagappan24, fortman25}, nonlinearities  \cite{boyi25}, and exceptional points \cite{hoerner24}, might find applications in a broader market, as the uncrypted information is directly accessible by naked eye. Further advances might leverage on the grey cavity matched CDC system to implement physically unclonable functions (PUFs), that are currently under intense scrutiny in the framework of photonics \cite{yang23, ferraro23, lin24responsive, klausen2025}. Moreover, optical nonlinearities could be included in the system, to enhance the phase sensitivity or to introduce an additional encryption parameter \cite{li18ol, alaee20, suwunnarat22}. Finally, recent acquisitions about coherent perfect absorbers like thermal noise and quantum-level effects \cite{vetlugin2019coherent, lai2024room, li21natphot, douglas24} might be the object of future integration within the proposed CPA-based cryptographic method. 

At the core of our proposal, we exploited the property that a thin film, whose conductivity is real and matched to the adjacent dielectric, operates as a broadband and lossy antireflection coating only for the waves traveling from the dielectric to the free-space. Ultimately, this results in an apparently grey cavity that becomes colored when observed in coherent absorption. We demostrated that this principle is easily implemented in the visible spectral range with a sample realized by means of ordinary nanofabrication techniques (silicon nitride membrane coated with gold and chromium thin films). However, by identifying appropriate materials (or metamaterals) that enforce the conductance matching property, the reported phenomenon can be extended seamlessly to the whole electromagnetic spectrum, likely enabling a number of impactful applications.

\section*{Methods}

\subsection*{Sample fabrication}
The fabrication procedure adopted to realize the multiple-window sample analyzed in Fig.~3 of main text is illustrated in Extended Data Figure 1. A commercial silicon wafer coated on both sides with $\approx 300\ \nm$ of low-stress, non-stoichiometric silicon nitride undergoes first multiple steps of processing on the front side. Each step consists of a sequence of optical lithography (Durham Magneto Optics ML3 system; resist S1818) and reactive-ion etching (RIE) for a calibrated amount of time in order to thin down the SiN layer up to desired thickness. RIE is performed in a SISTEC chamber with a $\mathrm{CF}_4/\mathrm{O}_2$ mixture, $40/2$ sccm, with a process pressure of $\approx 10^{-1}\ \mathrm{mbar}$; RF power was 40 W and bias $\approx -200\ \mathrm{V}$. After procesing the front side, the sample is processed on the back side through a single lithography/RIE step to define apertures in the SiN layer. Subsequently, wet etching in potassium hydroxide (30:70 w/w in DI water) heated at $80\ ^{\circ}\mathrm{C}$ is performed to create free-standing membranes. Finally, metal evaporation (EM Tecno Service thermal evaporator, base vacuum $5\cdot 10^{-6}\ \mathrm{mbar}$) is performed sequentially on front and back sides of the sample. The resulting sample cross section is illustrated in the lower-left part of Extended Data Figure 1a. Here, three different regions can be identified. Region 1 is opaque due to the bulk Si wafer; Region 3 is the desired multilayer; Region 2 is instead a transition region. For both Regions 1 and 2, the optical appearance from the sample front side is determined by the multilayer interference of the SiN/Si stack, which leads to quite strongly saturated colors with hues determined by the SiN thicness. This aspect is visible in Extended Data Figure 1b, which is the same as main Fig.~3c except for the absence of the dark grey frame. As described in the main text, and as visible here, the dark grey frame serves as a common reference background to properly identify the colors of Regions 3 without being misled by the color of Regions 1 and 2. We introduced the frame only in the reflection photograph, since both Regions 1 and 2 are opaque in transmission (due to the thick silicon substrate), and hence appear black in the transmission photograph (main Fig.~3b).
The ``smiley'' sample is fabricated following a similar procedure; here, however, the SiN regions with different thickness lie within a single $5\times5\ \mathrm{mm}^2$ sized membrane. We included corner markers made with a thick metallic film to ease the image processing procedure.

\subsection*{Optical setup}
The setup for single beam reflectance and transmittance measurement is based on a fiber coupled broadband lamp light source (DH-2000-BAL, Ocean Optics). The incident light beam was focused onto the sample to produce a circular spot with a diameter of about $250\ \micron$. Samples were positioned on a holder that is mounted on a rotation stage (PR01, Thorlabs), enabling precise control of the angle of incidence. Reflected or transmitted light was collected by a lens and coupled to an optical fiber (core diameter $1500\ \micron$, numerical aperture 0.39) connected to a monochromator for spectral analysis (Flame, Ocean Optics).

To measure the double beam absorption on the multi-frame sample (data in main Fig.~3g-i), we used the setup schematized in Extended Data Figure 2a. A He-Ne laser (nominal power 30 mW) is polarized and split in two branches; in one of these a liquid crystal (LC) phase shifter provides a phase delay tunable by approximately one wavelength. The beams are then directed onto the sample by means of two 50:50 beam splitters, and focalized through 100 mm focal length plano-convex lenses. The spot size is approximately $80\ \um$ in diameter. The beams outgoing from the sample are collected through two silicon detectors (approx.\ 8 mm sensitive area). The LC phase shifter and the detectors are operated through a custom-made electronic board interfaced to a PC through an Arduino DUE controller.

Extended Data Figure 2b represents the modified setup employed to measure the ``smiley'' sample (data in main Fig.~4c and Suppl.~Sect.~VIII. Here, the beams are no more focalized, rather, a beam expander is employed to obtain a $\approx\ 3\ \mm$ FWHM spot size. The beam outgoing from the sample are directed through a system of mirrors and a converging lens (focal length 15 cm) onto a CCD camera in two spatially separated positions. The LC is driven through the aforementioned Arduino-based system, while the CCD is read through the THORLABS native PC interface. This system lacks a trigger; however, analyzing the recorded frames, it is straightforward to recognize the time interval when the linear optical phase ramp is executed. Importantly, since a single camera is used, the coherent absorption videos from either sides of the sample are synchronized with each other, making it possible to perform the relative phase measurement discussed in the main text and in Suppl.~Sect.~VIII.

\section*{Acknowledgements}
Dr.~Raffaele Colombelli (CNRS, Paris Saclay), Dr.~Ruggero Verre (Chalmers University of Technology, Gothenborg) and Dr.~Francesco Riboli (CNR-INO, Firenze) are gratefully acknowledged for fruitful discussions and suggestions. CAMGRAPHIC (Pisa) is gratefully acknowledged for access to their ellipsometric facility. This work has been partially funded by the Royal Society (UK) under the grant IEC/R2/212012.

\section*{Author contributions} G.E.L.~and G.C.~performed the coherent absorption measurements and data analysis. L.L.~and A.C.~performed the transmission/reflection spectroscopic measurements. G.V.~and A.G.~contributed to the coherent absorption setup and data acquisition system. A.P.~and S.Z.~contributed to the sample fabrication. S.A.R.H., G.C.L.R., and A.T.~provided fundamental insights into the physics of the system under investigation. The work has been conceived and coordinated by S.Z., who also wrote the manuscript. All the authors have contributed to manuscript preparation and data review. 

\section*{Author current addresses}
Giulio Carotta: \textit{Institute of Physics, Ecole polytechnique fédérale de Lausanne (EPFL), CH-1015, Lausanne, Switzerland.}

Giacomo Venturi: \textit{Photonics Initiative, Advanced Science Research Center at the Graduate Center of the City University of New York, New York, NY 10031, USA}.

%
\newpage


\begin{figure}
    \centering
    \includegraphics[width=12 cm]{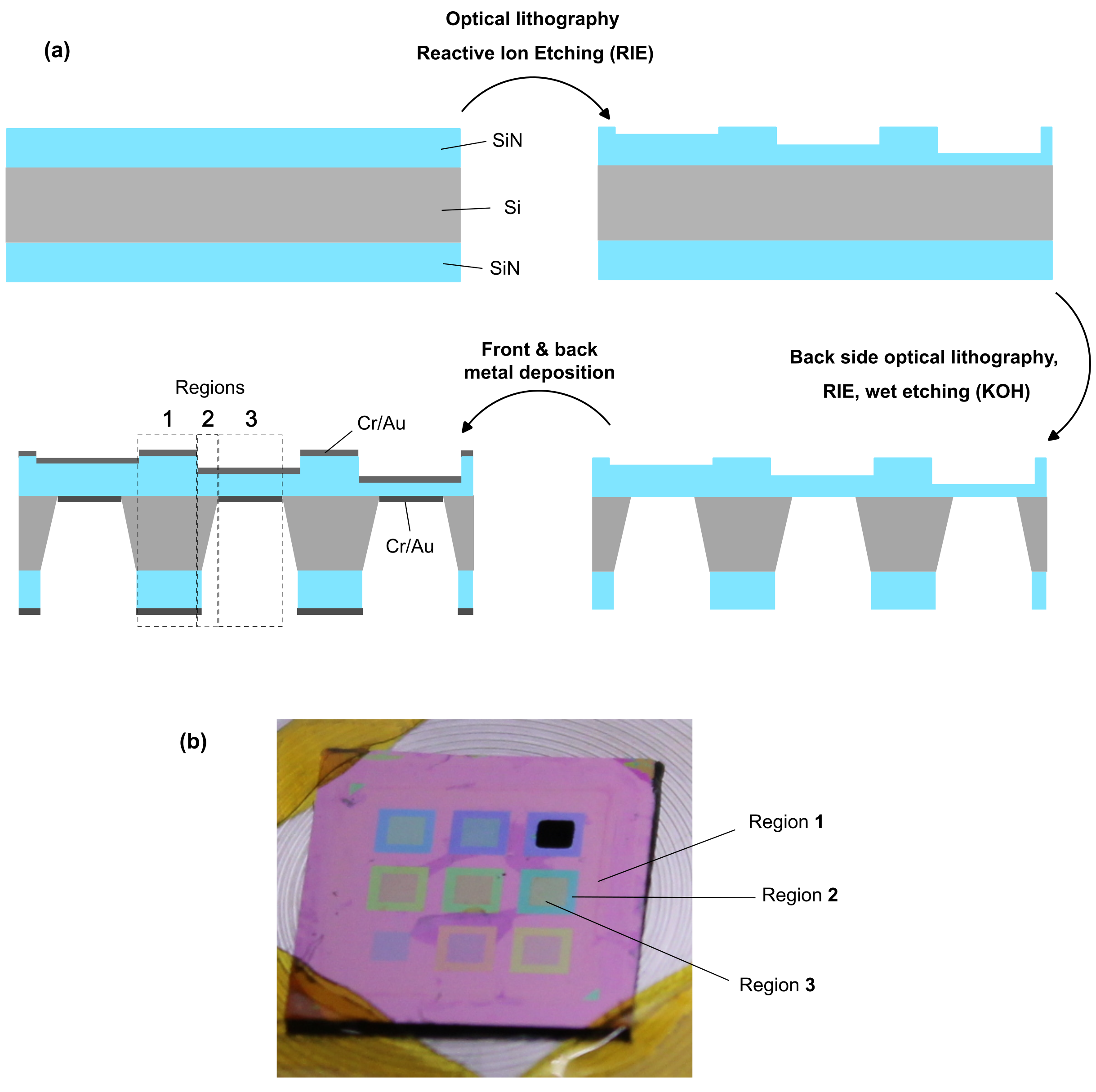}
    \caption*{\textbf{Extended Data Figure 1.} Schematic of the sample fabrication procedure and photograph of the fabricated sample.}
    \label{fig:s:fabrication}
\end{figure}

\clearpage

\begin{figure}
    \centering
    \includegraphics[width=12 cm]{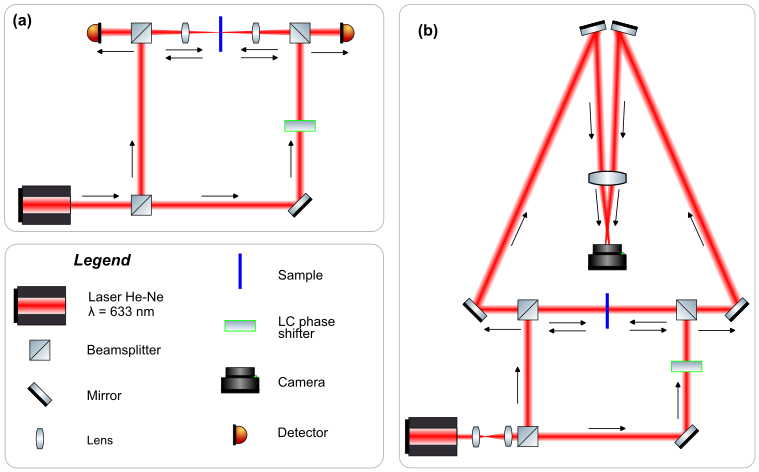}
    \caption*{\textbf{Extended Data Figure 2.} Schematic of the experimental setups used to perform the coherent absorption measurements.}
    \label{fig:s:setup}
\end{figure}


\clearpage

\renewcommand{\thepage}{S\arabic{page}} 
\setcounter{page}{1}                   
\onecolumngrid

\section*{Supplementary Materials}

\section*{I - Reflection and transmission coefficients of the conductor-dielectric-conductor (``CDC'') cavity}
\label{sect:s:mim}
\begin{figure}[h]]
    \centering
    \includegraphics[width=0.5\linewidth]{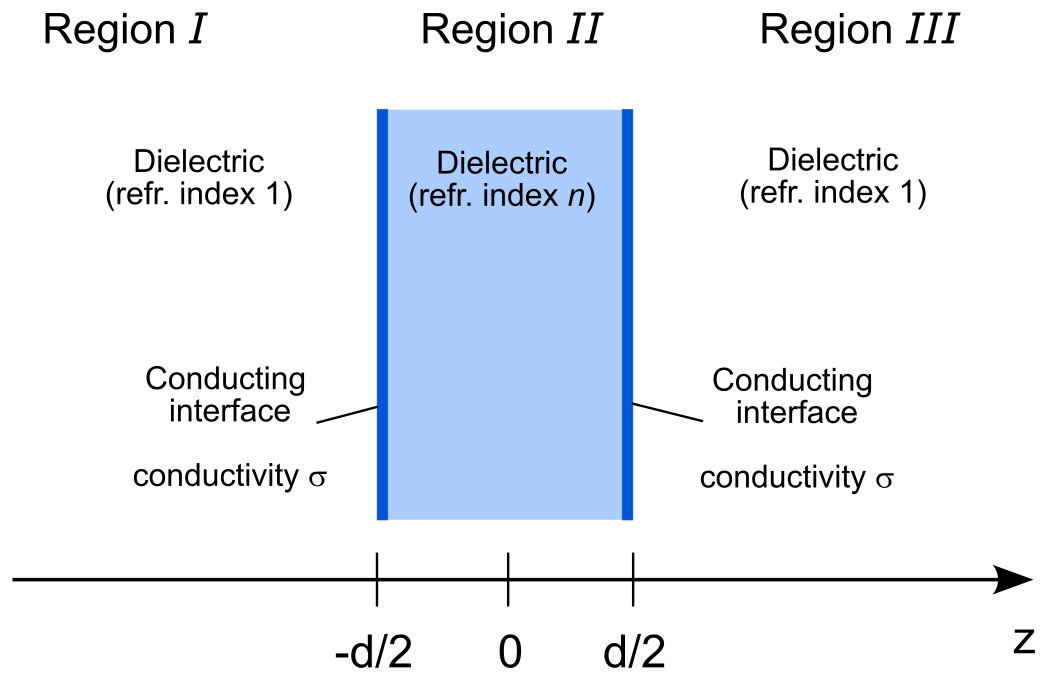}
    \caption{Schematic of the conductor-dielectric-conductor (CDC) cavity.}
    \label{fig:s:schematicsMIM}
\end{figure}

Let's consider a slab of material with real refractive index $n$ and thickness $d$ sandwiched between two surfaces with conductance $\sigma$ (in general complex), as described in Fig.~\ref{fig:s:schematicsMIM}, and calculate its reflection and transmission coefficients. 
The calculation itself is a standard propagation of electromagnetic waves normal to the interfaces. The fields in the different regions of the system can be expressed as:
\begin{align}
    \mathbf{E_I}(z)&=\mathbf{E_I^+}e^{ik_0(z+\frac{d}{2})}+\mathbf{E_I^-}e^{-ik_0(z+\frac{d}{2})}\\
    \mathbf{H_I}(z)&=\frac{1}{Z_0}(\mathbf{E_I^+}e^{ik_0(z+\frac{d}{2})}-\mathbf{E_I^-}e^{-ik_0(z+\frac{d}{2})})\nonumber\\
    \nonumber\\
    \mathbf{E_{II}}(z)&=\mathbf{E_{II}^+}e^{ikz}+\mathbf{E_{II}^-}e^{-ikz}\\
    \mathbf{H_{II}}(z)&=\frac{n}{Z_0}(\mathbf{E_{II}^+}e^{ikz}-\mathbf{E_{II}^-}e^{-ikz})\nonumber\\
    \nonumber\\
    \mathbf{E_{III}}(z)&=\mathbf{E_{III}^+}e^{ik_0(z-\frac{d}{2})}+\mathbf{E_{III}^-}e^{-ik_0(z-\frac{d}{2})}\\
    \mathbf{H_{III}}(z)&=\frac{1}{Z_0}(\mathbf{E_{III}^+}e^{ik_0(z-\frac{d}{2})}-\mathbf{E_{III}^-}e^{-ik_0(z-\frac{d}{2})})\nonumber
\end{align}
where $k_0=2\pi/\lambda_0$, $k=nk_0$ and we denoted with "$+$" the fields propagating in the $+z$ direction. The fields can be of any fixed linear polarization, as the materials and the interfaces are isotropic.
At the interfaces $z=\pm\ d/2$, the electric field is continuous (there is no charge accumulation at the infinitely thin conductor); while the magnetic field has a jump dictated by the surface current density $\mathbf{j} = \sigma \mathbf{E}$:
\begin{align}
    \mathbf{E_I}(\text{\tiny$-\frac{d}{2}$})=\mathbf{E_{II}}({\text{\tiny$-\frac{d}{2}$}})\quad &\Rightarrow\quad\mathbf{E_I^+}+\mathbf{E_I^-}= \mathbf{E_{II}^+}\beta^*+\mathbf{E_{II}^-}\beta \label{eq:MIM field match 1-2}\\
    \mathbf{H_I}(\text{\tiny$-\frac{d}{2}$})=\mathbf{H_{II}}({\text{\tiny$-\frac{d}{2}$}})+\sigma\mathbf{E_I}(\text{\tiny$-\frac{d}{2}$})\quad &\Rightarrow\nonumber\\
    \Rightarrow\quad\frac{1}{Z_0}(\mathbf{E_I^+}-\mathbf{E_I^-})&=\frac{n}{Z_0}(\mathbf{E_{II}^+}\beta^*-\mathbf{E_{II}^-}\beta)+\sigma(\mathbf{E_I^+}+\mathbf{E_I^-})\\
    \nonumber\\
    \mathbf{E_{II}}(\text{\tiny$\frac{d}{2}$})=\mathbf{E_{III}}({\text{\tiny$\frac{d}{2}$}})\quad &\Rightarrow\quad\mathbf{E_{III}^+}+\mathbf{E_{III}^-}= \mathbf{E_{II}^+}\beta+\mathbf{E_{II}^-}\beta^*\\
    \mathbf{H_{II}}(\text{\tiny$\frac{d}{2}$})=\mathbf{H_{III}}({\text{\tiny$\frac{d}{2}$}})+\sigma\mathbf{E_{III}}(\text{\tiny$\frac{d}{2}$})\quad &\Rightarrow\nonumber\\
    \Rightarrow\quad\frac{1}{Z_0}(\mathbf{E_{III}^+}-\mathbf{E_{III}^-})&=\frac{n}{Z_0}(\mathbf{E_{II}^+}\beta-\mathbf{E_{II}^-}\beta^*)-\sigma(\mathbf{E_{III}^+}+\mathbf{E_{III}^-})\label{eq:MIM field match 2-3}
\end{align}
where $\beta=e^{ik\frac{d}{2}}$. \\To obtain the reflection and transmission coefficients, we want to express the outgoing fields $\mathbf{E_I^-}$ and $\mathbf{E_{III}^+}$ as functions of the incoming ones  $\mathbf{E_I^+},\  \mathbf{E_{III}^-}$. To do this, we combine equations (\ref{eq:MIM field match 1-2}) to (\ref{eq:MIM field match 2-3}) to eliminate the terms $\mathbf{E_{II}^\pm}$, obtaining:
\begin{align}
    \frac{(1+a-b)}{\beta^*}\mathbf{E_I^+}+\frac{(1-a-b)}{\beta^*}\mathbf{E_I^-}&=\frac{(1+a+b)}{\beta}\mathbf{E_{III}^+}+\frac{(1-a+b)}{\beta}\mathbf{E_{III}^-}\label{eq:MIM conto1}\\
    \nonumber\\
    \frac{(1-a+b)}{\beta}\mathbf{E_I^+}+\frac{(1+a+b)}{\beta}\mathbf{E_I^-}&=\frac{(1-a-b)}{\beta^*}\mathbf{E_{III}^+}+\frac{(1+a-b)}{\beta^*}\mathbf{E_{III}^-}\label{eq:MIM conto2}
\end{align}
where $a=1/n$ and $b=\sigma Z_0/n$
By combining equations (\ref{eq:MIM conto1}) and (\ref{eq:MIM conto2}), the outgoing fields can be separated as:

\begin{multline}
\mathbf{E_I^-}=\frac{(1-a+b)(1-a-b)-(1+a-b)(1+a+b)}{\mathcal{D}}\mathbf{E_{III}^-}\\-\frac{(1+a-b)(1-a-b)\beta^2-(1-a+b)(1+a+b)\beta^{*2}}{\mathcal{D}}\mathbf{E_
{I}^+}\label{MIM out E 1}
\end{multline} 
\begin{multline}
    \mathbf{E_{III}^+}=\frac{(1+a-b)(1-a-b)\beta^2-(1-a+b)(1+a+b)\beta^{*2}}{\mathcal{-D}}\mathbf{E_{III}^-}\\+\frac{-(1-a+b)(1-a-b)+(1+a-b)(1+a+b)}{\mathcal{-D}}\mathbf{E_
{I}^+}\label{MIM out E 3}
\end{multline}
\begin{gather}
     \mathcal{D}=(1-a-b)^2\beta^2-(1+a+b)^2\beta^{*2}
\end{gather}
Using the ordinary definition of reflection and transmission coefficients and reintroducing the explicit forms of $a$, $b$ and $\beta$, we can obtain the following expressions:
\begin{gather}
    \frac{\mathbf{E_{I}^-}}{\mathbf{E_{I}^+}} \equiv r_{I\rightarrow I}=-\frac{(n+1-\sigma Z_0)(n-1-\sigma Z_0)e^{ikd}-(n-1+\sigma Z_0)(n+1+\sigma Z_0)e^{-ikd}}{(n-1-\sigma Z_0)^2e^{ikd}-(n+1+\sigma Z_0)^2e^{-ikd}}\label{MIM REFL gen}\\
    \nonumber\\
    \frac{\mathbf{E_{III}^+}}{\mathbf{E_{I}^+}} \equiv t_{I\rightarrow III}=\frac{(n-1+\sigma Z_0)(n-1-\sigma Z_0)-(n+1-\sigma Z_0)(n+1+\sigma Z_0)}{(n-1-\sigma Z_0)^2e^{ikd}-(n+1+\sigma Z_0)^2e^{-ikd}}.\label{MIM TRAS gen}
\end{gather}
Similar derivation lead to $r_{III\rightarrow III}$ and $t_{III\rightarrow I}$, which are respectively equal to $r_{I\rightarrow I}$ and $t_{I\rightarrow III}$, complying with the system symmetry. One can hence simply refer to $r$ and $t$ as in Eqs.~(2) and (3) of the main text.

\section*{II - Coherent absorption in a symmetric two-port system}
\label{sect:s:cpa}
We here derive the general formulas for coherent absorption in a symmetric two-port system. With reference to main Fig.~1c, incoming and outgoing waves are related by the following scattering matrix relation:
\begin{equation}
\mqty(a^{I}_{\out} \\ a^{II}_{\out})=S\mqty(a^{I}_{\in} \\ a^{II}_{\in})=\mqty(r&t\\t&r) \mqty(a^{I}_{\in} \\ a^{II}_{\in})
\label{eq:s:Smatrix}
\end{equation}
When the system is illuminated with equal-intensity, dephased beams
\[
\mqty(a^{I}_{\in} \\ a^{II}_{\in}) = \frac{1}{\sqrt{2}}\mqty(1\\e^{i\phi}),
\]
the outgoing waves carry intensities
\begin{align}
\label{eq:s:iIiIIout}
I^{I}_{\out} = |a^{I}_{\out}|^2 & = \frac{1}{2}|r+t\, e^{i \phi}|^2 \nonumber \\ \nonumber
             & = \frac{1}{2}\left(|r|^2+|t|^2+2|rt| \cos (\phi-\psi) \right)\\ \nonumber \\ \nonumber
I^{II}_{\out} = |a^{II}_{\out}|^2 & = \frac{1}{2}|t+r\, e^{i \phi}|^2 \\ 
             & = \frac{1}{2}\left(|r|^2+|t|^2+2|rt| \cos (\phi+\psi) \right)
\end{align}
where $\psi = \arg(r/t)$. Notice that both $I^{I}_{\out}$ and $I^{II}_{\out}$ are modulated with a peak-to-peak amplitude of $2|rt|$. 
When performing a coherent absorption measurement, the absolute input phase difference $\phi$ is rarely known, as it would require stabilizing the interferometer path length and the sample position with nanometer accuracy. Rather, one performs a \textit{phase sweep}, for instance through a delay line (moving mirror) or through a tunable liquid crystal cell. In this way, the $\phi$ sweep is known accurately except for an undefined offset. In this framework, one can easily access experimentally the \textit{phase difference} between the $I^{I,II}_{\out}$ sinusoids, which is $(-\psi) - \psi = -2\psi \equiv \Delta$. The phase $\Delta$ can be identified by simply inspecting the relative position of the maxima of the measured $I^{I,II}_{\out}$ sinusoids, or better by a fitting procedure as detailed in Suppl.~Sect.~\ref{sect:s:CoherentMeas}.

Once known the individual output beam intensities, the total output intensity is determined as
\begin{equation}
I_{\out} = I^{I}_{\out} + I^{II}_{\out} = |r|^2+|t|^2+2|rt| \cos \psi \, \cos \phi.
\end{equation}

One can then define the \textit{double beam absorptance} by $A_{db} = 1-I_{\out}$; evidently this is a $\phi$-dependent quantity. Recalling that $A_{sb}= 1-|r|^2-|t|^2$ is the single-beam absorptance, one can write
\begin{equation}
\label{eq:s:adb}
A_{db} = A_{sb}-2|rt| \cos \psi \, \cos \phi.
\end{equation}
It follows straightforwardly that the minimum and maximum double beam absorptance (with respect to $\phi$) are given by $A_{db,m} = A_{sb}- 2|rt \cos \psi|$ and $A_{db,M} = A_{sb}+ 2|rt \cos \psi|$. Moreover, it shall be noticed that averaging $A_{db}$ with respect to $\phi$ gives exactly $A_{sb}$. 

\subsection*{A - Alternative derivation based on singular value decomposition}
As known\footnote{Ge, Li, and Liang, Feng. ``Contrasting eigenvalue and singular-value spectra for lasing and antilasing in a PT-symmetric periodic structure.'' Physical Review A \textbf{95}, 013813 (2017).}, the minimum (maximum) double beam absorptance (respectively, $A_{db,m}$ and $A_{db,M}$) are related to the singular values of the $S$-matrix via the relation
\begin{align}
A_{db,m} & = 1-s_>^2 \nonumber \\
A_{db,M} & = 1-s_<^2 
\label{eq:s:Adb_sigma}
\end{align}
where $s_>$ ($s_<$) are the maximum (minimum) singular values\footnote{To identify the singular values, we use here the letter $s$ instead of the more common letter $\sigma$ to avoid confusion with the conductivity. }.
In the present case (Eq. S\ref{eq:s:Smatrix}), the singular values are 
\begin{align}
s_1 = |r+t| \nonumber \\ 
s_2 = |r-t| 
\label{eq:s:sigma12}
\end{align}
where we are using subscripts $\{1,2\}$ rather than $\{ <,>\}$ since it is not in general known which among $|r\pm t|$ is the largest number.

\section*{III - Coherent absorption in the matched conductor-dielectric-conductor cavity}
Using the results of Suppl.~Sect.~\ref{sect:s:mim} specialized for the matched case $\sigma = \sigma_M = (n-1)/Z_0$ (i.e.~ $r = (1-n)/n$, $t = e^{ikd}/n$) in the formulas of Suppl.~Sect.~\ref{sect:s:cpa}, one gets 
\begin{align}
\Delta &= -2\psi = 2kd, \\ \nonumber \\
A_{sb} &= 1-\frac{(1-n)^2}{n^2} - \frac{1}{n^2} = 2 \frac{n-1}{n^2}, \label{eq:s:Asb}\\ \nonumber \\
2|rt| &= 2\frac{n-1}{n^2}. 
\end{align}
and eventually
\begin{equation}
A_{db} = A_{sb} \left( 1 - \cos(kd)\cos\phi \right)
\label{eq:s:Adb_mim}
\end{equation}
which is Eq.~(1) in the main text.
Two non-general facts occur here: first, the phases $\psi$ and $\Delta$ are determined solely by the product $kd$, while in the non-matched CDC model they are complicated functions of $\sigma$ and $kd$. In more general two-port systems, $\psi$ and $\Delta$ are a function of geometric parameters, permittivities, wavelength etc.
Moreover, here $2|rt| = A_{sb}$; this allows to factorize $A_{sb}$, which again is not a general fact.
An obvious minimization (maximization) with respect to $\phi$ gives
\begin{align}
A_{db,m} = A_{sb} \left( 1 - |\cos(kd)| \right) \label{eq:Adb_m_matched} \\
A_{db,M} = A_{sb} \left( 1 + |\cos(kd)| \right) \label{eq:Adb_M_matched} 
\end{align}

\subsection*{ A - Alternative derivation based on singular value decomposition}
Using Eq.~S\ref{eq:s:sigma12}, $r = (1-n)/n$, and $t = e^{ikd}/n$, one gets straightforwardly
\[
s_{1,2}^2 = \frac{2-2n+n^2}{n^2} \mp \frac{2(n-1)}{n^2}\cos(kd)
\]
where the minus sign applies to $s_1$. Recalling Eq.~S\ref{eq:s:Adb_sigma} and Eq.~S\ref{eq:s:Asb}, one obtains
\begin{align*}
A_{db,1} & = 1-s_1^2 = A_{sb}(1+\cos(kd)) \\
A_{db,2} & = 1-s_2^2 = A_{sb}(1-\cos(kd))
\end{align*}
which match either Eqs.~S\ref{eq:Adb_m_matched} and S\ref{eq:Adb_M_matched}.

\section*{IV - The matched dielectric-conductor-dielectric interface}
\begin{figure}
    \centering
    \includegraphics[width=0.45\linewidth]{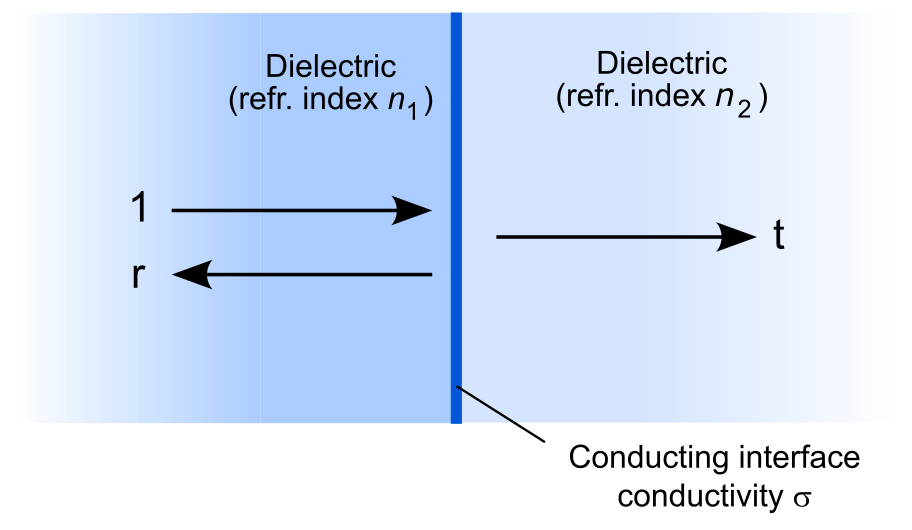}
    \caption{Schematic of the dielectric cavity sandwiched between conducting sheets.}
    \label{fig:s:schematicsMI}
\end{figure}

Consider two dielectric regions of real refractive indices $n_1$ and $n_2$ separated by a conducting interface of complex conductivity $\sigma$ (Fig.~\ref{fig:s:schematicsMI}), and a wave at normal incidence. 
Straightforward calcultions similar to those of Suppl.~Sect.~\ref{sect:s:mim} give
\begin{eqnarray}
r = \frac{n_1-n_2-\sigma Z_0}{n_1+n_2+\sigma Z_0}\\ \nonumber \\
t = \frac{2n_1}{n_1+n_2+\sigma Z_0}.
\end{eqnarray}
These formulas generalize both the ordinary Fresnel coefficient at normal incidence (for $\sigma = 0$), and the reflection/transmission formulas for a thin conducting sheet in vacuum, usually employed by the radiofrequency and the graphene community (for $n_1 = n_2 = 1$). A peculiar, usually overlooked feature of these formulas is that one can set $r=0$ by just matching the interface conductance to the appropriate real value, i.e.~using $\sigma Z_0 = n_1-n_2$. The caveat is that \textit{only the reflection seen from the high-index side} can be easily zeroed, as otherwise a real and negative $\sigma$ (i.e.~gain) would be needed. Moreover, if such a matched interface is realized, it is necessarily lossy, due to the finite real part of the conductivity. The single beam absorption can be calculated as $A_{sb} = 1-|r|^2-n_1/n_2|t|^2$ (where the $n_1/n_2$ factor stems from the bulk wave impedance mismatch), and reads
\[
A = \frac{n_1-n_2}{n_1} = \frac{n_1/n_2-1}{n_1/n_2}
\]
i.e.~an absorption that \textit{grows} with the refractive index contrast $n_1/n_2$.

\section*{V - Effect of real metals}
\label{sect:s:realmetals}
\begin{figure}
    \centering
    \includegraphics[width=0.8\linewidth]{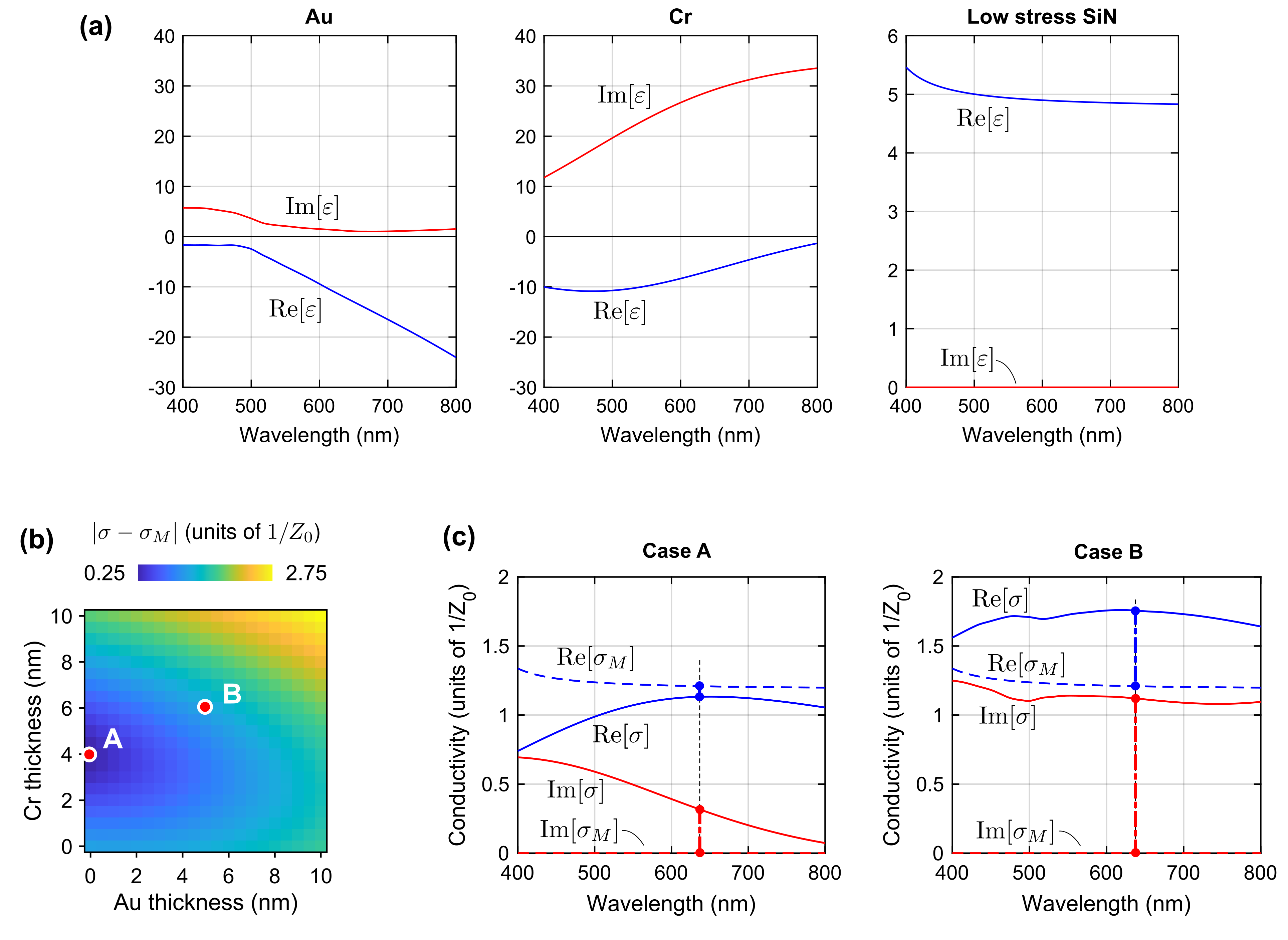}
    \caption{Relative permittivity of the materials used in the sample, and deviation of the thin-film conductivity from the ideally matched conductivity $\sigma_M$.}
    \label{fig:s:conductivities}
\end{figure}

In Fig.~\ref{fig:s:conductivities}a we report the wavelength-dependent relative permittivities of gold, chromium and silicon nitride, where we have employed tabulated values for the metals and data obtained from ellipsometry for the silicon nitride. Based on these data, and employing 
\[
\sigma = -i \left[ d_{\mathrm{Cr}} (\eps_{\mathrm{Cr}}-1) + d_{\mathrm{Au}} (\eps_{\mathrm{Au}}-1) \right] \eps_0 2 \pi c/\lambda
\] 
we determined the deviation of the effective metal bilayer conductivity from the ideal one, $\sigma_M = (n_{\mathrm{SiN}}-1)/Z_0$. The deviation, calculated at $\lambda = 633\ \nm$, is reported in Fig.~\ref{fig:s:conductivities}b, as a function of the individual metal layer thicknesses. As discussed in the main text, the ideal case is $d_{\mathrm{Cr}} = 4\ \nm$, $d_{\mathrm{Au}} = 0\ \nm$ (Case A); this is reasonable since $\re (\eps_{\mathrm{Cr}}-1)$ is much smaller than $\im (\eps_{\mathrm{Cr}})$, making the resulting $\sigma$ closer to a pure real value. However, as discussed in the main text, to avoid islanding effects and to protect the bare Cr surface, we adopted the choice $d_{\mathrm{Cr}} = 6\ \nm$, $d_{\mathrm{Au}} = 5\ \nm$ (Case B). In Fig.~\ref{fig:s:conductivities}c we report the wavelength dependent conductivity of both Case A and Case B, together with the ideal $\sigma_M$.

\section*{VI - Coherent absorption single-point data processing}
\label{sect:s:CoherentMeas}
To process the coherent absorption data collected from the multiple-membrane sample using the setup in Extended Data Figure 2a, we fitted the experimental data according to the following expression:
\begin{equation}
\label{eq:s:fitting}
I_{\out, \mathrm{exp}}^{\alpha}(n)  = p_1^{\alpha} + p_2^{\alpha} \cos( 2\pi p_3^{\alpha}\, n/N + p_4^{\alpha})
\end{equation}
where $p_{1}^{\alpha} \ldots p_4^{\alpha}$ are free fitting parameters, $\alpha = I,II,\mathrm{or}\ I+II$, and $n$ is an index that runs from 1 to $N$, being $N$ the total number of samples contained in a full phase sweep. Since the phase sweeper is physically constructed in order to obtain a sweep of approximately one wavelength, it is expected that $p_3 \approx 1$, which was indeed the case. The experimental data, together with the fitted sinusoids, are reported in main Fig.~3g.
From that figure it can be noticed that $p_1^{I}$ is very close to $p_1^{II}$, i.e.~the average signals on either sides are equal, as expected from the theory (see Eq.~S\ref{eq:s:iIiIIout}). Similarly, also the sinusoid amplitudes were similar, as confirmed from $p_2^{I}$ and $p_2^{II}$ being very close with each other. Instead, $p_4^{I}$ and $p_4^{II}$ are different. What however matters is the difference $p_4^{I} - p_4^{II}$, that should be compared with $-2 \psi \equiv \Delta$ (main Fig.~3i). 

\clearpage
\section*{VII - Single-beam spectra in the non-perfectly-matched cavity}
\label{sect:s:singlebeam}
\begin{figure}[h]
    \centering
    \includegraphics[width=0.8\linewidth]{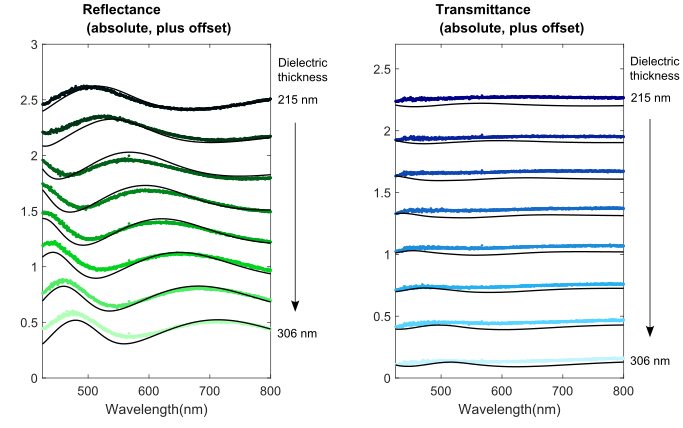}
    \caption{Single-beam spectra for variable-thickness CDC cavities. Colored traces are experimental data; black traces are the result of a multilayer model. A constant offset has been applied to better visualize the trend of the spectral features.}
    \label{fig:s:singlebeam}
\end{figure}

We report in Fig.~\ref{fig:s:singlebeam} as colored traces the same spectra plotted in main Fig.~3d-e, applying however a constant offset to better visualize the trend of the spectral features. Moreover, we here plot the theoretical spectra determined by a multilayer model\footnote{The ``PPML'' toolbox for MATLAB has been employed, \url{https://it.mathworks.com/matlabcentral/fileexchange/55401-ppml-periodically-patterned-multi-layer}}, where all the layers are simulated with their dispersive permittivities (see Suppl.~Sect.~\ref{sect:s:realmetals}). 

\section*{VIII - Coherent absorption image processing}
\label{sect:s:CoherentImage}
\begin{figure}[h]
    \centering
    \includegraphics[width=0.95\linewidth]{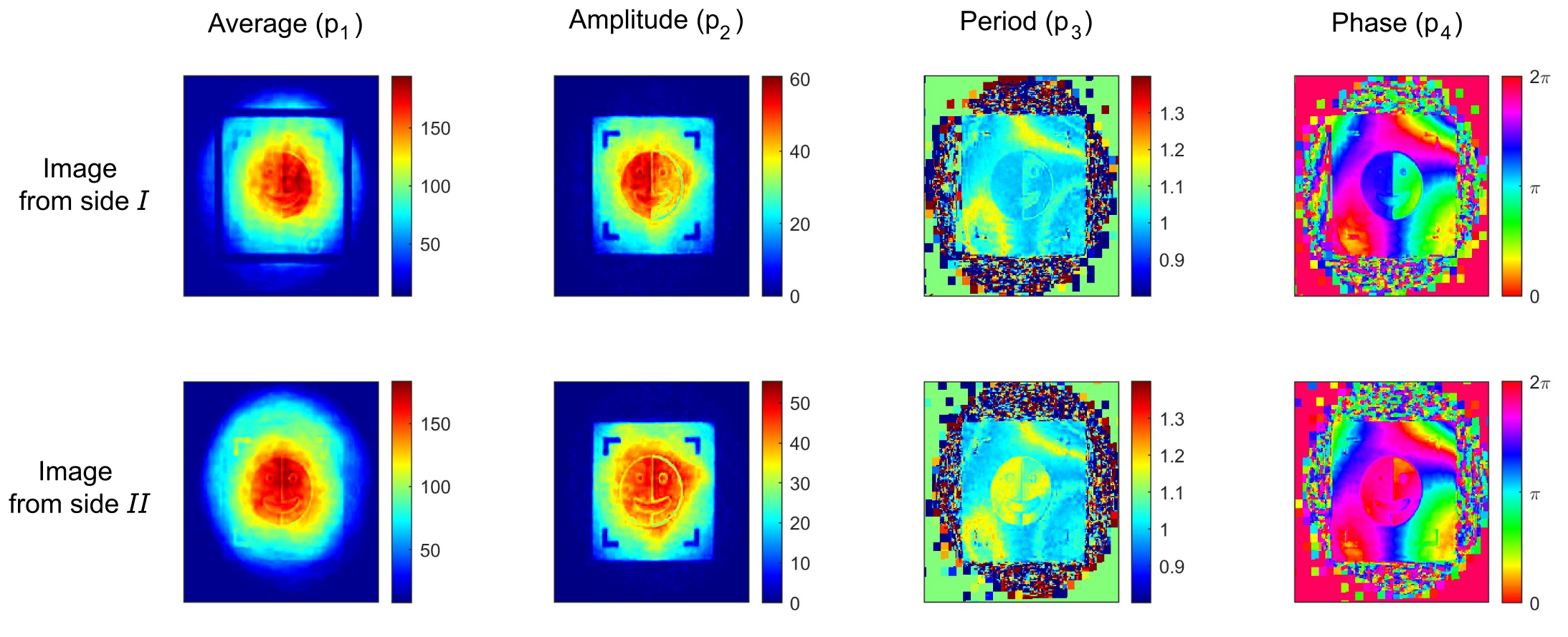}
    \caption{Images collected from the two sides of the ``smiley'' sample while coherent illumination is employed.}
    \label{fig:s:faccinaCPA}
\end{figure}

To process the coherent absorption image, we basically performed the fitting of Eq.~S\ref{eq:s:fitting} pixel-by-pixel on the sequence of frames collected with the CCD of Extended Data Figure 2b. Prior to the fitting, the images are cropped and overlapped using an affine transformation with the aid of the corner markers. We hence obtain images of parameters $p_1^{\alpha}\ldots p_4^{\alpha}$, reported in Fig.~\ref{fig:s:faccinaCPA} for $\alpha = I, II$. Here, several features can be seen:
\begin{itemize}
\item The average ($p_1$) keeps trace of the laser Gaussian spot, while the ``smiley'' is almost invisible. This is consistent with the fact that in theory $p_1$ should be equal to $\frac{1}{2}\left( |r|^2 + |t|^2 \right)$, which is independent on the thickness when the conductivity has the matched value;
\item The amplitude ($p_2$), again, keeps trace of the laser Gaussian spot, plus some hints of the ``smiley'' profile. This is almost consistent  with the fact that in theory $p_2$ should be equal to $|rt|$, which is independent on the thickness when the conductivity has the matched value;
\item The normalized period $p_3$ is close to unity, and is well-defined only for the pixels where the intensity modulation was large enough (i.e.~inside the window, where interference takes place);

\item The phase $p_4$ shows evidence of the ``smiley'' plus a kind of structured background. This background originates from the imperfect overlap of the incoming wavefronts, and is strongly dependent upon the beam alignment. Such misalignment has the effect of changing the reference phase for each pixel; nonetheless, the reference phase is constant throughout the experiment, and once the phase difference $p_4^{I} - p_4^{II}$ is considered (main Fig.~4c) the ``smiley'' structure emerges, as expected from the dependence of $\Delta$ upon the dielectric thickness. 
\end{itemize}

\end{document}